\documentclass[sigconf]{acmart}
\usepackage{graphicx}
\usepackage{colortbl}
\usepackage{etoolbox}
\usepackage{multirow}
\usepackage{makecell}
\usepackage{color}
\usepackage[utf8]{inputenc}
\usepackage{xspace}
\usepackage{colortbl}
\usepackage{pifont}
\usepackage{tabularx}

\AtBeginDocument{%
  }

\setcopyright{acmcopyright}
\copyrightyear{2023}
\acmYear{2023}
\acmDOI{XXXXXXX.XXXXXXX}

\acmConference[Conference '23]{}{2023}{arxiv}
\acmPrice{15.00}
\acmISBN{978-1-4503-XXXX-X/18/06}

\begin{document}

\title{AI Friends: Designing Creative Coding Assistants for Families}

\author{Stefania Druga}
\orcid{0000-0002-5475-8437}
\affiliation{%
  \institution{University of Washington}
  \city{Seattle}
  \state{Washington}
  \country{USA}
  \postcode{98195}}
\email{st3f@uw.edu}
\author{Amy J. Ko}
\affiliation{%
 \institution{University of Washington}
  \city{Seattle}
  \state{Washington}
  \country{USA}
  \postcode{98195}}
\email{ajko@uw.edu}

\renewcommand{\shortauthors}{Druga, et al.}

\begin{teaserfigure}
\centering
\includegraphics[width=\columnwidth]{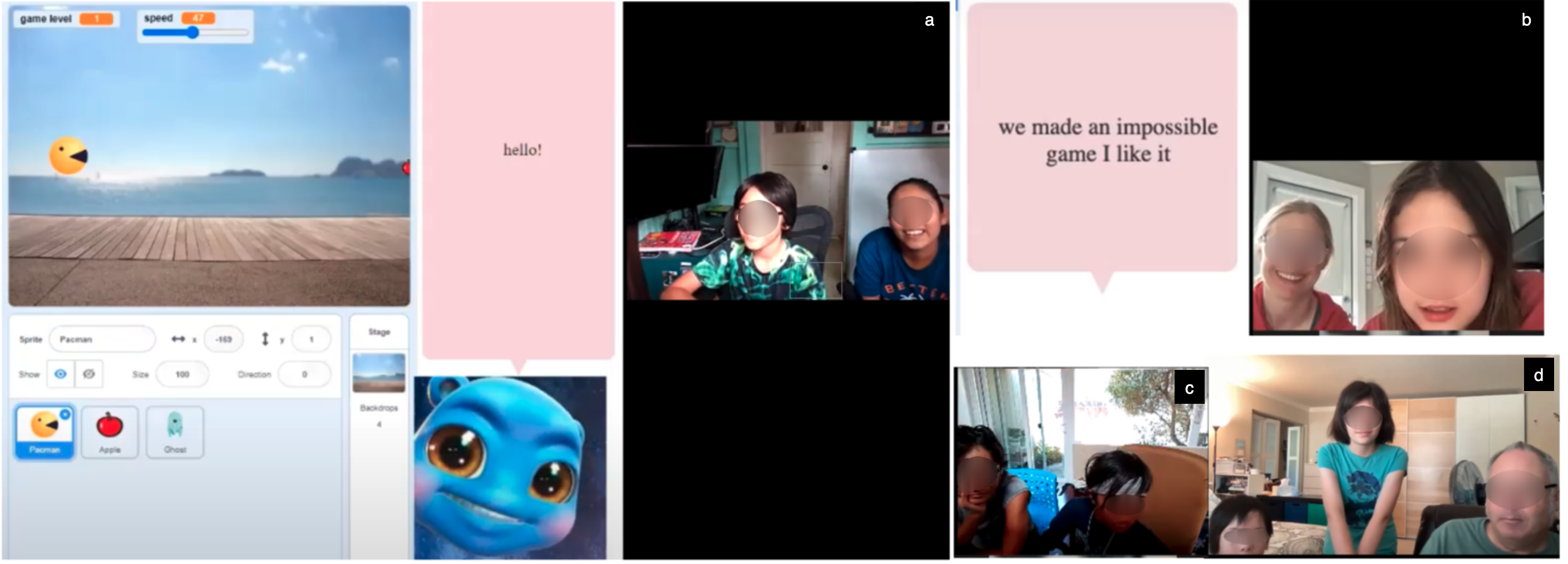}
\caption{Examples of joint family interaction during the study: (a) F1 getting acquainted with the AI Friend, (b) F5's reaction to a joke from an AI Friend, (c) F8 mom and son playing a multiplayer Pacman game they just programmed, and (d) F4 dad and two daughters brainstorming game ideas with the AI Friend.}
\label{fig:fig_joint_interaction}
\end{teaserfigure}
\begin{abstract}
What role can AI play in supporting and constraining creative coding by families? To investigate these questions, we built a Wizard-of-Oz platform to help families engage in creative coding in partnership with a researcher-operated AI Friend. We designed a 3-week series of programming activities with ten children (7 to 12 years old) and nine parents. Using a creative self-efficacy lens, we observe that: (1) families found it easier to generate game ideas when prompted with questions by AI Friend, (2) parents played a unique role in guiding children in more complex programming tasks when the AI Friend failed to help, and (3) children were more encouraged to write code for novel ideas using the AI friend's help. These findings suggest that AI-supported platforms should highlight unique family-AI interactions focused on children’s agency and creative self-efficacy.

\end{abstract}




\maketitle

\section{Introduction}
Computing education has a long and eclectic history of engaging youth in creative coding \cite{kelleher2005lowering}. Since the early 1960s, ever-larger youth communities have been making with code. Scratch, for example, has engaged more than 120 million youth in creating with code \cite{ScratchS68:online}, and more recent programmable platforms such as Minecraft and Roblox have engaged youth in curating environments for play. Moreover, creative coding ``blurs the distinction between art and design and science and engineering''\cite{levin2021code}, encompassing diverse youth interests such as generative art, embedded computing, audio editing, performative live programming, and countless others.

Youth make with code not to learn to code \textit{per se}, but to express themselves, connect, and play \cite{kelleher2007storytelling, kafai2013connected}, as well as to learn and reason \cite{resnick1993reasoning}. Therefore, coding by youth is a form of computing education that also provides additional benefits to youth development and well-being. Further, past research shows that youth need to engage in this type of creative learning more \cite {ito2013connected}. Youth create with friends and peer mentors, in camps, at school, after school, at home, and online \cite{ko2018informal, roque_im_2016, druga_how_2021}. 

Youth also create with family \cite{druga20214as,druga2021idc, druga2022family, banerjee2018empowering}.These inter-generational contexts for learning and expression afford many unique opportunities. For example, the Reggio Emilia educational philosophy, which centers families as a site of reflective learning, emphasizes listening, documentation, and critical reflection \cite{Fawcett2004-zg}. This pedagogy is based on four core principles: (1) creative values are children's strength and power, and attending to these reinforces the value of listening, (2) creative relationships are attentive and respectful, (3) creative environments are physical and emotional, and (4) behavior and dispositions matter for holistic support of learning and creative thinking. Similarly, Joint-Media Engagement (JME) notions help explore how families come together to learn, shifting roles and responsibilities in guiding creative activities \cite{stevens_new_2011}. This demonstrates how families are another context for learning and, when computing concepts are engaged, for computing education.

While there has been much exploration of tools to support creative expression and learning with code \cite{kelleher2005lowering}, support for creativity itself is more limited. Most of these support tools attempt to reduce errors \cite{Cooper2003}, facilitate debugging \cite{Ko2004-whyline}, or otherwise support software engineering activities \cite{ko2011state}. These productivity supports are different from creativity supports \cite{Resnick2005-fx}, which might encourage exploration, diversity of expression, multiple paths to the same expression, and collaboration. Unfortunately, far fewer creative coding support tools exist compared to productivity tools, and there are especially few that facilitate family learning \cite{zan2022neural}.

Supporting complex creative processes such as ideation and collaboration is challenging. These are, by definition, non-linear, iterative, and unpredictable activities, making them less amenable to carefully structured tool support. Large language models, however, offer new possibilities. For example, AI-powered code assistants such as GitHub's Copilot \cite{GitHubCo82:online} and Replit's Ghostwriter \cite{Ghostwri56:online} suggest a future in which the creativity inherent to professional programming is directly supported by intelligent assistants responding to a diversity of prompts and contexts.

The emerging feasibility of creative coding support for professionals suggests the following research question: \textit{How might children and parents engage in collaborative creative coding supported by AI?} This question is critical not only to imagining creative coding support tools that empower children to learn but also to do so in a way that reflects the increasingly AI-assisted future of computing and computing education.

A systematic and informed approach to this question requires observing children and parents engaging in creative coding supported by an intelligent agent, \textit{before} building the agent. Our observations aim to generate a collection of rich and ``thick'' \cite{geertz2008thick} descriptions of how youth might collaborate with AI for creative coding. By generating these insights on diverse creative coding contexts, including both open-ended \cite{Tsur2018-tp} and closed-ended projects \cite{Lee2014-gidget}, we seek to inform the design of AI-driven creativity support but also surface its limitations and caveats.

To this end, we employed a Wizard-of-Oz (WoZ) methodology, creating the illusion of intelligent support, which we called an ``AI Friend'', and observing families interact with it to support their creative learning. Widely used in HCI research \cite{dow2005_woz}, including in feasibility studies of AI-assisted tasks \cite{jung2014_woz_arduino,kahn2016_woz_robot, ruan2020_woz_math}, this approach allowed us to flexibly explore how the presence of an AI assistant might change family learning dynamics without needing to build an agent.
We designed our AI Friend to be attentive, respectful, and friendly, following prior work in Human-Robot Interaction (HRI) on family-AI interaction, which shows a strong effect of personality cues \cite{ostrowski2021long}. We then ran a 3-week longitudinal study with 19 participants (10 children aged 7-12 and 9 parents in 8 families) and observed families' creative coding practices and interaction with the AI Friend trying to help them. 

Our investigation makes three contributions to the understanding of AI-supported creative coding:

\begin{enumerate}
    \item We provide insight into how families engage in collaborative creative coding with an AI friend. 
    \item We document the unique benefits of families' joint engagement with AI supports. 
    \item We use a theoretical model of creative self-efficacy and explain how it relates to developing AI-supported creative coding in families.
\end{enumerate}

Our study found that it was easier for families to generate game ideas when prompted with questions by AI Friend and that parents guided children in more complex programming tasks when the AI Friend could not help. These findings suggest that AI-supported coding platforms should highlight unique family-AI interactions and support children’s agency and creative self-efficacy.

\section{Background}
This section presents prior work on family joint-creative coding, an overview of relevant tools for facilitating creativity, and the theory of creative-self efficacy and its uses in prior family studies. 

\subsection{Family Joint-Engagement in Creative Coding}
We frame our investigation of family learning from the perspective of \textit{Joint-Media Engagement} (JME), which Stevens et al. defined as “spontaneous and designed experiences of people using media together” ~\cite{stevens_new_2011}. Their analysis focused on the six ideals of productive JME: (1) mutual engagement, (2) dialogic inquiry, (3) co-creation, (4) boundary-crossing, (5) intention to develop, and (6) focus on content, not control~\cite{stevens_new_2011}.

JME has been studied extensively in the context of computing education. For example, prior work has found that parents, peers, and caregivers can play a dynamic role in youth learning. They can act as facilitators or guides \cite{barron_parents_2009}, learners, or lead youth to see themselves as experts \cite{michelson2021parenting, chi_aifam2022}. 
Families can also bridge formal learning at school and informal student-driven learning outside of school \cite{mitra2006youth}. Other studies have demonstrated that parental experience in technology fields significantly impacts how they support their children’s learning \cite{disalvo2014they}. Family-oriented programs, such as Family Creative Learning (FCL) \cite{roque2016m, roque2015engaging}, are essential for families lacking “preparatory privilege” \cite{margolis2010stuck} to get involved in their children’s creative coding activities.

Research on family use and perception of coding has revealed that parents’ primary concern about supporting their children’s computing literacy is their limited programming knowledge~\cite{yu_considering_2020}. Designers have explored text-free programming platforms to support parents better, finding that families can create together successfully \cite{banerjee2018empowering, govind2020engaging}. Further understanding AI programming in family contexts may uncover new opportunities to link youth interests in AI with interest-driven programming \cite{cunningham2020purpose}, family relationships \cite{musick2021gaming}, and formal computing education \cite{barnes2007game2learn}.

\subsection{Tools for Facilitating Creativity}
Our work also builds upon perspectives at the intersection of creativity and media. At the highest level, our study concludes that every child has immense natural talents \cite{Gardner2013-imagination} and innate creative potential \cite{Vygotsky2004-jt}. These premises raise a central question of how to design new learning opportunities and tools for creative thinking that allow families to flourish in an era of rising technology consumption.

One aspect of this question is the media itself. What young people create today heavily depends on the tools and materials used. From Froebel gifts, \cite{wiggin1895froebel} to LEGO Mindstorms \cite{klassner2003lego}, and creative learning tools such as Scratch \cite{maloney2010scratch}, notable efforts have been made to foster creative learning and coding for youth. These initiatives flourished primarily outside of traditional educational institutions, leveraging two critical aspects of creativity for children: allowing them to tinker, construct, debug, test, and modify ideas and encouraging them to collaborate in person or digital communities. The success of these projects has also driven change in the way creative thinking and coding are taught in schools, with more initiatives focused on project-based learning and coding. 

Nevertheless, questions remain as to how best to balance structure and agency in programming for youth \cite{Brennan2015-hf}.
A growing body of work suggests that technology-enabled tools could effectively scaffold parent-child activities; however, most have focused on supporting remote parent-child communication. 
For example, numerous projects have analyzed how technology-enabled systems can provide a virtual space for parents and children to interact \cite{judge2010sharing, Sun2016, yarosh2008supporting}. Other studies have explored how to support remote parent-child activities, such as facilitating gameplay \cite{Follmer2010, hunter2014waazam} or reading together \cite{raffle2010family}. Recent work on parent-child interactions in co-located contexts has studied multi-touch tabletop applications \cite{xiao2012supporting}, sensor-based exergames \cite{saksono2015spaceship}, and technology-enhanced storytelling activities \cite{teepe2017technology, cingel2017parents}. Although this work informs design, little prior work has considered coding specifically.

Some work has considered creativity support more directly. A recent systematic literature review study sheds light on Creative Support Technologies (CSTs) \cite{Frich2019-zj}. The study found six significant categories of support in the creative process: pre-ideation, idea generation, evaluation, implementation, iteration, and reflection. As Frich et al. point out, many CSTs are disconnected from the creators’ daily practices \cite{Frich2019-zj}. In the context of computing education, youth want their learning to be authentic \cite{Shaffer1999-at}. Authenticity in creative coding could involve providing the proper media support, like in the case of the danceON project \cite{Payne2021-danceon}, and the opportunity to work on microworlds with curated programming activities, such as fashion or music \cite{Tsur2018-tp}.
These examples suggest different ways children and families might work with their CSTs and engage in creative coding with intelligent systems \cite{Karimi2018-eu}.

\subsection{Creative Self-Efficacy and Families}
Our study also builds upon notions of \textit{creative self-efficacy theory (CSE)} \cite{Tierney2002-ib}.
Derived from Bandura’s more general concept of self-efficacy \cite{Bandura1977-nn}, creative self-efficacy concerns confidence in one’s ability to be creative. It asserts that an individual’s beliefs about their creativity impact their willingness to attempt the creative task, their level of effort, and the duration of their persistence when faced with difficulty \cite{Tierney2002-ib}.

Prior work on creative self-efficacy had explored its development and performance over time, observing that when individuals perceived recognition for their creative performance and if their supervisor expected them to be creative, their creative self-efficacy improved with time. 
Furthermore, an increase in creative performance was linked to a higher degree of creative self-efficacy \cite{Tierney2002-ib}: when someone succeeds in a task and has a “mastery experience,” their self-efficacy regarding the task will increase; conversely, when someone has high self-efficacy for a task, they will accomplish it at a higher level than if they had a lower sense of self-efficacy \cite{Bandura1977-nn}. 

CSE has been used by creativity-support tools (CST) designers to focus on children’s and parents’ beliefs that they can successfully perform a specific creative process. 
For example, Mosaic, an online creative community, builds creative self-efficacy by sharing the design process for creative work rather than showcasing finished projects \cite{Kim2017-mosaic}. 
Parallel prototyping in creative work leads to better design results and increased self-efficacy~\cite{Dow2011-nd}. 
In addition, the Creativity Project utilized CSE theory when designing a mobile game to engage youth in different kinds of creative thinking and behavior at a science exhibit; results indicated that playing the game contributed to increased creative self-efficacy for participants \cite{atwood2019creative}.

Some prior work has examined CSE in families, finding that children’s creative self-efficacy can be either positively or negatively influenced by parent–child relationships in after-school program activities \cite{liang2020exploring}. Gralewski et al. found that parental child acceptance and autonomy support were weakly but positively related to children’s creative self-efficacy and creative personal identity \cite{gralewski2020parenting}. Tang et al. found that parental support and creative self-efficacy significantly predicted student creative self-efficacy in studies on parental influences on student general and Science, Technology, Engineering, and Mathematics (STEM) creative ideation behaviors \cite{tang2022parents}. These findings inform our conceptions of CSE in the context of our joint-family creative coding study design and analysis.

\section{Method}
Building upon principles of Joint-Media Engagement (JME) principles, gaps in Creativity Support Tools (CST) for creative coding, and conceptions of creative self-efficacy (CSE), our study asked \textit{How might children and parents engage in collaborative creative coding supported by an AI Friend?}. To answer this, we designed a Wizard-of-Oz (WoZ) AI Friend to examine how AI Friends might need to be designed to promote JME in a way that promotes creative self-efficacy. 

Eight families participated in three online study sessions, each lasting 30-40 minutes. We conducted two sessions of games programming for different games with AI Friends and one final interview. During the programming sessions, families interacted with an AI Friend, controlled remotely by the researcher. The AI Friend provided creative prompts, coding debugging, and ideas to support families. At the end of each session, families engaged in a final interview to offer feedback on the AI Friend and suggest new features or designs.

\subsection{Study Participants}
We recruited 8 families (10 children aged 7-12 and 9 parents) from 6 different US states for our study. Families had a wide range of socio-economic backgrounds and spoke 5 languages other than English (see demographics in Table \ref{tab:woz_demogs}). Families also declared various levels of exposure to AI technologies and programming in the screening survey. In addition, each family member completed an intake questionnaire and described their programming experience. The study took place via video conference.

\begin{table*}[]
\begin{tabular}{clll}
\textbf{Family ID} & \textbf{Parent(s)} & \textbf{Language(s)} & \textbf{Child(ren) and Age(s)} \\ \hline
F1 & Mom (S.), Dad (J.) & English, Spanish & Son, 7 (G.) \\
\rowcolor[HTML]{EFEFEF} 
F2 & Mom (T.) & English, Spanish & Daughter, 12 (H.) \\
F3 & Dad (J.) & English & Son, 11 (G.) \\
\rowcolor[HTML]{EFEFEF} 
F4 & Dad (D.) & English, Mandarin & \begin{tabular}[c]{@{}l@{}}Daughter, 12 (A.) \\ Daughter, 10 (M.)\end{tabular} \\
F5 & Mom (C.) & English & Daughter, 10 (K.) \\
\rowcolor[HTML]{EFEFEF} 
F6 & Mom (M.) & \begin{tabular}[c]{@{}l@{}}French, Cantonese, \\ Mandarin\end{tabular} & \begin{tabular}[c]{@{}l@{}}Son, 11 (Z.) \\ Daughter, 7 (K.)\end{tabular} \\
F7 & Dad (M.) & English, Japanese & Son, 10 (M.) \\
\rowcolor[HTML]{EFEFEF} 
F8 & Mom (L.) & English, Tagalog & Son, 12 (S.)
\end{tabular}
\caption{Families participating in the study.}
\label{tab:woz_demogs}
\end{table*}

\subsection{Study Sessions}

\textbf{Session 1: Modifying a Coding Micro-world.}
The first author introduced families to the CogniSynth platform and the AI Friend in this session. The AI Friend then guided them through the rest of the activity, presenting them with a list of 3 different micro-worlds \cite{Tsur2018-tp} (Fish Game, Drawing Game, and Pacman Game micro-worlds) and asking them to pick one and modify it to make it more fun. We decided only to give participants these three options to scope their potential explorations, building on prior work that shows constraints can be a source of creative inspiration \cite{caniels2015organizing}. Throughout the activity, the AI Friend provided encouragement, ideas, and reflection questions to the families.

\textbf{Session 2: Choosing Programming Patterns to Create a Game.} 
In this session, the AI Friend gave families the task of picking three programming patterns from a given collection and using them to create a game. Examples of programming patterns were provided, showing code scripts for creating different game events (e.g., firing objects or jumping over obstacles).

\textbf{Session 3: Final Interview.} 
During the final interviews, the families interacted with the researcher and reflected on their interactions with the AI Friend and their study experience. The researcher showed them video snippets from their prior sessions and asked them to describe what they did and when the agent was helpful. They were also shown screenshots from their study games and asked to identify moments when the AI Friend was helpful. Families were asked to give feedback on alternative user interface (UI) designs and describe how they would like to interact with the AI Friend to get ideas, debugging help, or encouragement. Finally, each family rated different attributes of the AI Friend on a scale. The researcher collected copies of all completed games and analyzed them for correctness, diversity of features, and uniqueness.

\subsection{Study Materials}
\subsubsection{The CogniSynth Platform} 
\begin{figure*}
\centering
\includegraphics[width=5.5in]{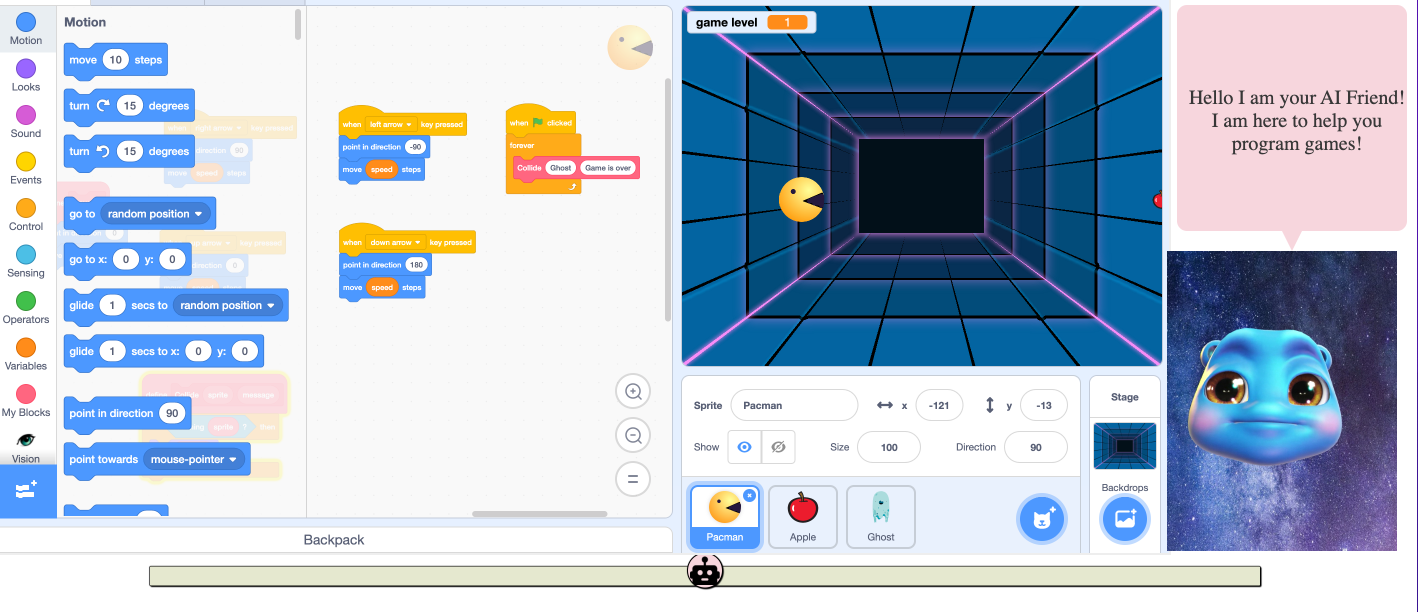}
\caption{The CogniSynth family view interface consisted of three components: (1) the Coding Blocks Library and Coding area, (2) the AI Friend Response area, and (3) the AI Friend Avatar window.}
\label{interface}
\end{figure*}

\begin{figure}
\centering
\includegraphics[width=2.5in]{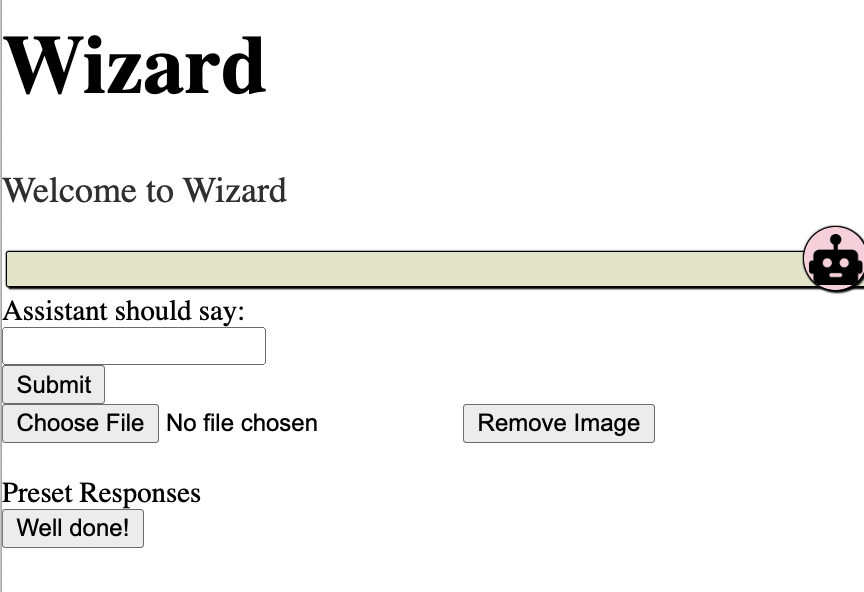}
\caption{The CogniSynth wizard interface consisted of three components: a Dialogue window, a Quick Reactions menu, and a list of prompts.}
\label{interface_woz}
\end{figure}

\begin{figure}
\centering
\includegraphics[width=2.7in]{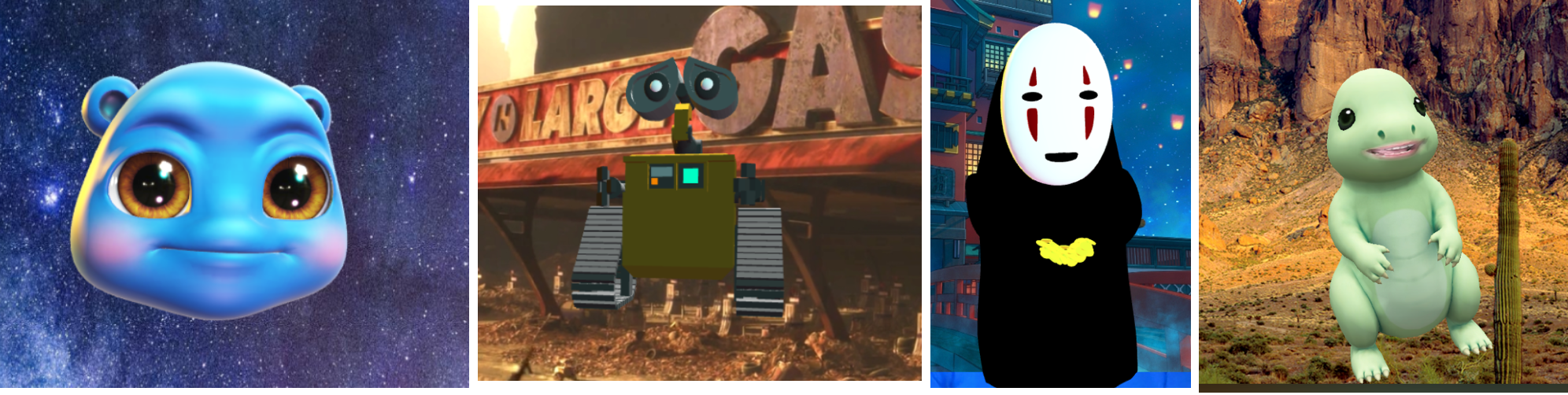}
\caption{The list of AI Friends included Water Bear, Wall-E, Maskman, and Dinosaur.}
\label{avatars}
\end{figure}

We designed and built the CogniSynth platform for this proposed study. It has two main views: family and wizard. The \textit{family view} consists of a \textit{Coding Blocks Library} and \textit{Coding} area, an \textit{AI Friend Response} area, and an \textit{AI Friend Avatar} window (see Figure \ref{interface}). The \textit{wizard view} enabled us to create the illusion of an intelligent assistant. It had a \textit{Dialogue} window, a \textit{Quick Reactions} menu, and a list of prompts. The researcher could write messages that were sent via the AI Friend in real-time or select a pre-written message or prompt via a keyboard shortcut. The researcher could also upload and send images in the \textit{Dialogue} window (see Figure \ref{interface_woz}). To allow access to the \textit{AI Friend Window} in the family view, the researcher first opened the wizard view and installed Snap Camera Studio. A custom Snap Camera filter was developed for each AI Friend, which tracks the researcher's face and expressions and maps them to control various 3D avatars (see Figure \ref{avatars}). 

\subsubsection{AI Friends}
For each AI Friend, we created a custom Snap Camera filter. This filter tracks the researcher's face and expressions and maps them to control various 3D avatars (see Figure \ref{avatars}). When no face is detected, a pre-set background is displayed. We created the filters using the Snap Studio SDK and open-source 3D models from thingiverse.com. Each AI Friend communicates with families on the screen using a text-based  dialog.

\subsubsection{Creative Micro-worlds}
To help children explore possibilities without being limited by a starter program, we created three themed micro-worlds: ```Drawing micro-world'', ``Pacman Game micro-world'' and ``Fish Game micro-world''. The ``Drawing micro-world'' allowed families to create a custom drawing program by selecting colors, stamps, and brushstroke effects. The ``Pacman Game micro-world'' enabled families to program a Pacman game with Pacman and Ghost characters they can move around on the screen. The ``Fish Game micro-world'' enabled families to program a game with big fishes that eat smaller fishes.

\begin{figure*}
    \centering
    \includegraphics[width=4.5in]{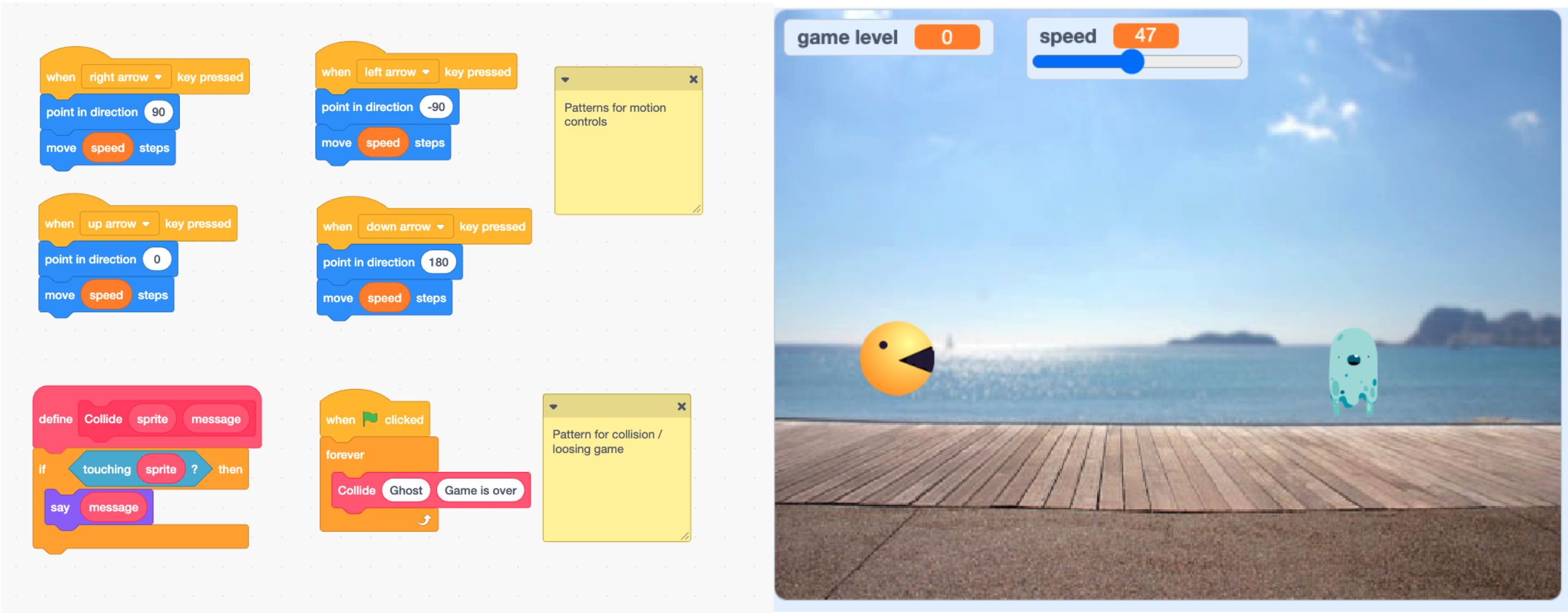}
    \caption{The Game micro-world. Available blocks (left) and the game controlled by the program (right).}
    \label{fig:game-micro-world}
\end{figure*}

We carefully designed micro-world consisting of just a few blocks to express various programs to engage families' creativity. For example, in the ``Pacman Game micro-world'' (Figure ~\ref{fig:game-micro-world}) the micro-world was made up of blocks that prompt a user for commands, control Pacman movement, animate Ghosts, and check the input for a condition. This small set of blocks can be combined to create various programs, allowing families to create many variations on the Pacman game or modify it into a new game. In our prior work, when we gave families such micro-worlds instead of the complete coding platform, they had an easier time constructing valid programs and imagining potential behaviors for game characters \cite{druga2022familiescode}.

\subsubsection{AI Friend Prompts}
Building on prior work on social robots supporting youth creativity in drawing, storytelling, and LEGO programming \cite{ali2021social}, we curated and adapted a list of creative prompts for this study. Depending on the family action, we selected any wizard view prompt (or typed other similar ones). The AI Friend then displayed the text in the family coding area. Table \ref{tab:aifriend_prompts} shows the prompts that the AI Friend could make. To devise these, we used free and creative writing methodologies for inspiration \cite{earnshaw2007handbook}, grouping them into reflective questions, creative prompts, and positive reinforcement. The AI Friend only performed these text prompts and did not engage in other forms of behavior. The AI Friend could also reply in writing to questions and prompts from participants.

\begin{table*}[]
\begin{tabular}{clc}
\textbf{Reflective Questions} &
  \multicolumn{1}{c}{\textbf{Creative Prompts}} &
  \textbf{Positive Reinforcement} \\ \hline
\rowcolor[HTML]{EFEFEF} 
\textit{\begin{tabular}[c]{@{}c@{}}Can you tell me why you \\ did that?\end{tabular}} &
  \textit{\begin{tabular}[c]{@{}l@{}}What are some other things you \\ can make your project do?\end{tabular}} &
  \textit{\begin{tabular}[c]{@{}c@{}}That is such a great idea! \\ Good job.\end{tabular}} \\
\textit{What will you do next?} &
  \textit{\begin{tabular}[c]{@{}l@{}}What else can you make the \\ character do in this situation?\end{tabular}} &
  \textit{\begin{tabular}[c]{@{}c@{}}You think of some really \\ cool rules for the game.\end{tabular}} \\
\rowcolor[HTML]{EFEFEF} 
\textit{What are you trying to make?} &
  \textit{\begin{tabular}[c]{@{}l@{}}Can you make it do \\ something else?\end{tabular}} &
  \textit{\begin{tabular}[c]{@{}c@{}}Well done. \\ That was so creative!\end{tabular}} \\
\textit{How are you going to do that?} &
  \textit{\begin{tabular}[c]{@{}l@{}}Let’s think of some fun uses \\ of the game.\end{tabular}} &
  \textit{\begin{tabular}[c]{@{}c@{}}I would not have thought \\ of that. Good going.\end{tabular}} \\
\rowcolor[HTML]{EFEFEF} 
\textit{\begin{tabular}[c]{@{}c@{}}What are the blocks \\ would you need for that?\end{tabular}} &
  \textit{\begin{tabular}[c]{@{}l@{}}Is there a better way to \\ program this event?\end{tabular}} &
  \textit{Great solution!} \\
\textit{\begin{tabular}[c]{@{}c@{}}Do you have any questions\\  about this script?\end{tabular}} &
  \textit{\begin{tabular}[c]{@{}l@{}}Let’s try to make an obstacle\\ for the character.\end{tabular}} &
  \textit{Oh, that is fun!} \\
\rowcolor[HTML]{EFEFEF} 
\textit{Is that the best way to do that?} &
  \textit{\begin{tabular}[c]{@{}l@{}}Let’s try to make the character \\ move when you press space\end{tabular}} &
  \textit{I love this animation!} \\
\textit{\begin{tabular}[c]{@{}c@{}}Let's read the code\\ and see what it does\end{tabular}} &
  \textit{\begin{tabular}[c]{@{}l@{}}Should we animate the \\ character?\end{tabular}} &
  \textit{Great character design!}
\end{tabular}
\caption{Examples of prompts used by the AI Friend.}
\label{tab:aifriend_prompts}
\end{table*}

\subsubsection{Programming Patterns}
We created a collection of \textit{eight programming patterns} that families could use as examples when programming their games in Session 2 of our study. Our patterns provided examples of game behaviors, such as throwing objects and animated motion.
We decided to include explicit patterns as a form of scaffolding, guided by prior work that demonstrates the importance of  \textit{programming plans}, small program fragments that achieve a goal, such as selecting values from a list that match specific criteria \cite{soloway1980problems}; prior work on children's game programming demonstrated that providing programming patterns and templates for different game types can facilitate computational literacy and expression \cite{Troiano2020-av, hsiao2018automated, Franklin2020-ah, kynigos2007half}.

\begin{figure*}
\centering
\includegraphics[width=6in]{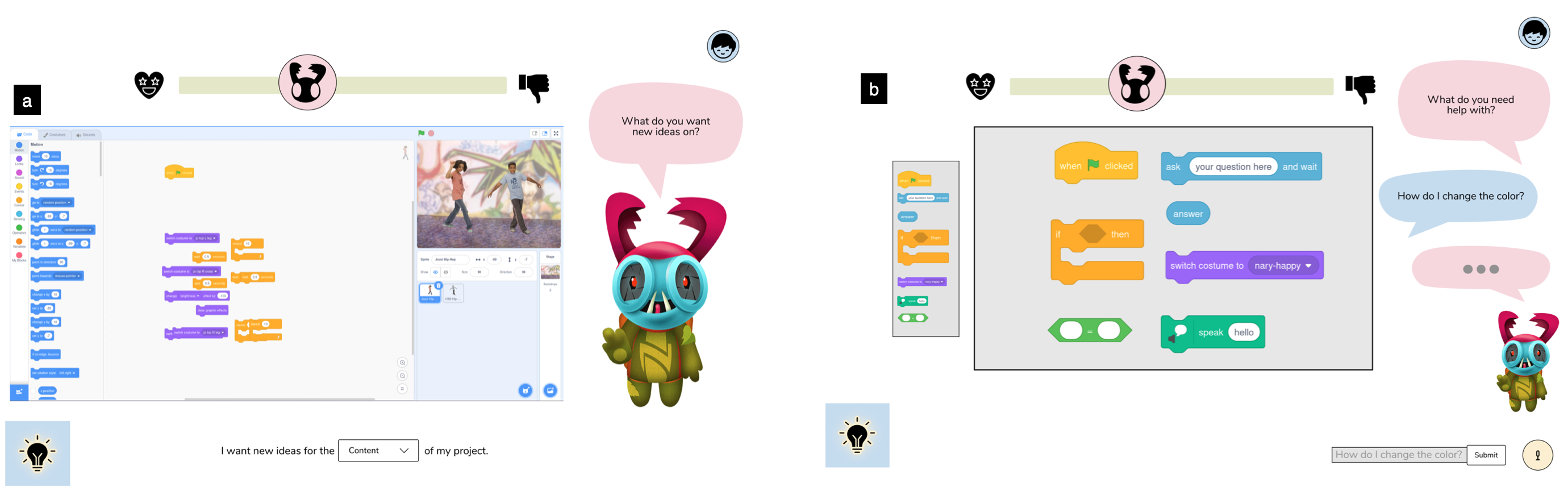}
\caption{Examples of platform UI mock-ups used for co-design: (a) options for eliciting specific ideas from the AI Friend, and (b) options for different chat modalities with AI Friend.}
\label{codesign}
\end{figure*}

\subsubsection{Co-Designing UI Mock-ups}
During the final study session, we provided user interface mock-ups to involve families in designing creative support behaviors for the AI Friend. This included different ways for it to express new ideas during coding and different modes of interaction with it (e.g., via text, audio, or images). In addition, families were asked to provide feedback on these scenarios and suggest their feature ideas or new ways to interact with existing features. Figure \ref{codesign} shows examples of the UI mock-ups used for co-design.

\subsubsection{AI Persona Sheet}
At the end of the study, families completed the AI Persona sheet (see Figure \ref{fig:ai_perception}). This sheet contained different potential characteristics of the AI Friend depicted on continua. Family members marked where the AI Friend fell on the scale for each characteristic and then discussed why they made those choices. The list of characteristics was adapted from a survey designed by Bartneck et al.\cite{bartneck2009measurement}; this instrument has been frequently used to measure children's anthropomorphism, animacy, perceptions of likeability, perceptions of intelligence, and perception of the safety of robots. Though the original instrument examines perceptions across 24 items, we adapted the items to focus on a subset of 9 characteristics. 

\subsection{Data Collection and Analysis}
We collected video recordings of all study sessions and in-situ activity feedback and reflections from the slider provided on the platform. We also recorded logs of all written support provided by the AI Friend. Both children and parents were encouraged to speak aloud \cite{charters2003use} during the programming activities. After the activities, family members were prompted to describe the interaction with the AI Friend in a final interview.

For our qualitative analysis, we transcribed the videos and noted comments on families’ body language and non-verbal interactions. We analyzed each transcript using a combination of etic codes developed from our CSE theoretical framework and emic codes that emerged from the interviews themselves \cite{miles1984drawing,patton1990qualitative}. After developing a final codebook (see Table.\ref{tab:aifriend_codes}), we coded all transcripts and then used this coding process to develop categories, which we conceptualized into broad themes \cite{braun2006using}. We also transcribed and analyzed family reflections and feedback from the final interviews and their co-design sheets for the AI Friends. To protect our participant identities, we chose to have the first author exclusively run interviews, clean transcripts, and analyze the data. Only the first author maintained access to participant identifiers and performed coding; the second author only saw themes from the analysis and selected quotations. Methodologically, this maximized coherence between data collection and analysis, in line with recent work~\cite{vaithilingam2022expectation}, but risked a greater potential for bias in our results. We felt it more important to take cautious steps forward than aim for objectivity. However, we emphasize that our goal was to inform future research on creative coding AI assistance design rather than verify the objectivity of our claims; the latter would require multiple coders and inter-rater reliability~\cite{hammer2014confusing}). 
Finally, we collected and analyzed all the projects created by participating families. Each project was analyzed for correctness, diversity of features, and uniqueness. We also traced the provenance of ideas to show what code came from the AI or parental suggestions.

\begin{table*}[]
\begin{tabular}{clc}
\textbf{Code} &
  \multicolumn{1}{c}{\textbf{Definition}} &
  \textbf{Example} \\ \hline
\rowcolor[HTML]{EFEFEF} 
\textit{\begin{tabular}[c]{@{}c@{}}Produce ideas\end{tabular}} &
  \begin{tabular}[c]{@{}l@{}}AI friend helps families come up\\ with ideas for their games\end{tabular} &
  \textit{\begin{tabular}[c]{@{}c@{}}"How can a ghost\\ go over the walls?"\end{tabular}} \\
\textit{\begin{tabular}[c]{@{}c@{}}Express ideas in code \end{tabular}} &
  \begin{tabular}[c]{@{}l@{}}AI friend helps families find \\ the right programming blocks\\ to express their games ideas\end{tabular} &
  \textit{\begin{tabular}[c]{@{}c@{}}"How can you \\ make the ball move\\ faster?"\end{tabular}} \\
\rowcolor[HTML]{EFEFEF} 
\textit{Debug} &
  \begin{tabular}[c]{@{}l@{}}AI friend helps with code \\ reading, debugging and testing\end{tabular} &
  \textit{\begin{tabular}[c]{@{}c@{}}"Let's test the \\ "shoot"script"\end{tabular}} \\
\textit{\begin{tabular}[c]{@{}c@{}}Elaborate on AI\end{tabular}} &
  \begin{tabular}[c]{@{}l@{}}Family members build on \\ suggestions from AI friend\end{tabular} &
  \textit{\begin{tabular}[c]{@{}c@{}}"Oh I like the \\ zombie ghost idea\\ let's make it green"\end{tabular}} \\
\rowcolor[HTML]{EFEFEF} 
\textit{\begin{tabular}[c]{@{}c@{}}Elaborate on family \end{tabular}} &
  \begin{tabular}[c]{@{}l@{}}AI friend makes suggestions\\ building on family ideas\end{tabular} &
  \textit{\begin{tabular}[c]{@{}c@{}}"Should the bear \\ say "Ouch" when\\ touching hedgehog?"\end{tabular}} \\
\textit{AI failure} &
  \begin{tabular}[c]{@{}l@{}}Instances when AI friend fails\\ to help or provide useful ideas\end{tabular} &
  \textit{\begin{tabular}[c]{@{}c@{}}"Maybe your mom\\ can help with clones?"\end{tabular}} \\
\rowcolor[HTML]{EFEFEF} 
\textit{\begin{tabular}[c]{@{}c@{}}Joint-engagement\\ support\end{tabular}} &
  \begin{tabular}[c]{@{}l@{}}Instances when AI friend supports\\ family joint-engagement\end{tabular} &
  \textit{\begin{tabular}[c]{@{}c@{}}"How about letting\\ your sister code the \\ taco animation?"\end{tabular}} \\
\textit{\begin{tabular}[c]{@{}c@{}}Creative coding\\ identity\end{tabular}} &
  \begin{tabular}[c]{@{}l@{}}AI friend helps kids develop \\ their creative coding identity\end{tabular} &
  \textit{\begin{tabular}[c]{@{}c@{}}"I love your dynamic \\ maze idea! So fun!"\end{tabular}}
\end{tabular}
\caption{Summary of final codes and definitions for family creative coding with an AI Friend.}
\label{tab:aifriend_codes}
\end{table*}

\section{Results: Collaborative Creative Coding with AI Support}
We divide the results into three sections. First, we overview what families created in each session, to give the reader a sense of what happened in the family's creative collaboration, and provide a high-level sense of the role of the AI friend in these collaborations. Then, we present a series of sections, each corresponding to the themes that emerged in \ref{tab:aifriend_codes}, discussing the role of the AI friend in each specific aspect of family interaction. Finally, we end with insights from the family co-design sessions.

\subsection{Session and Project Overview of Family-AI Friend Interactions}

\begin{figure*}[t]
\centering
\includegraphics[width=5.5in]{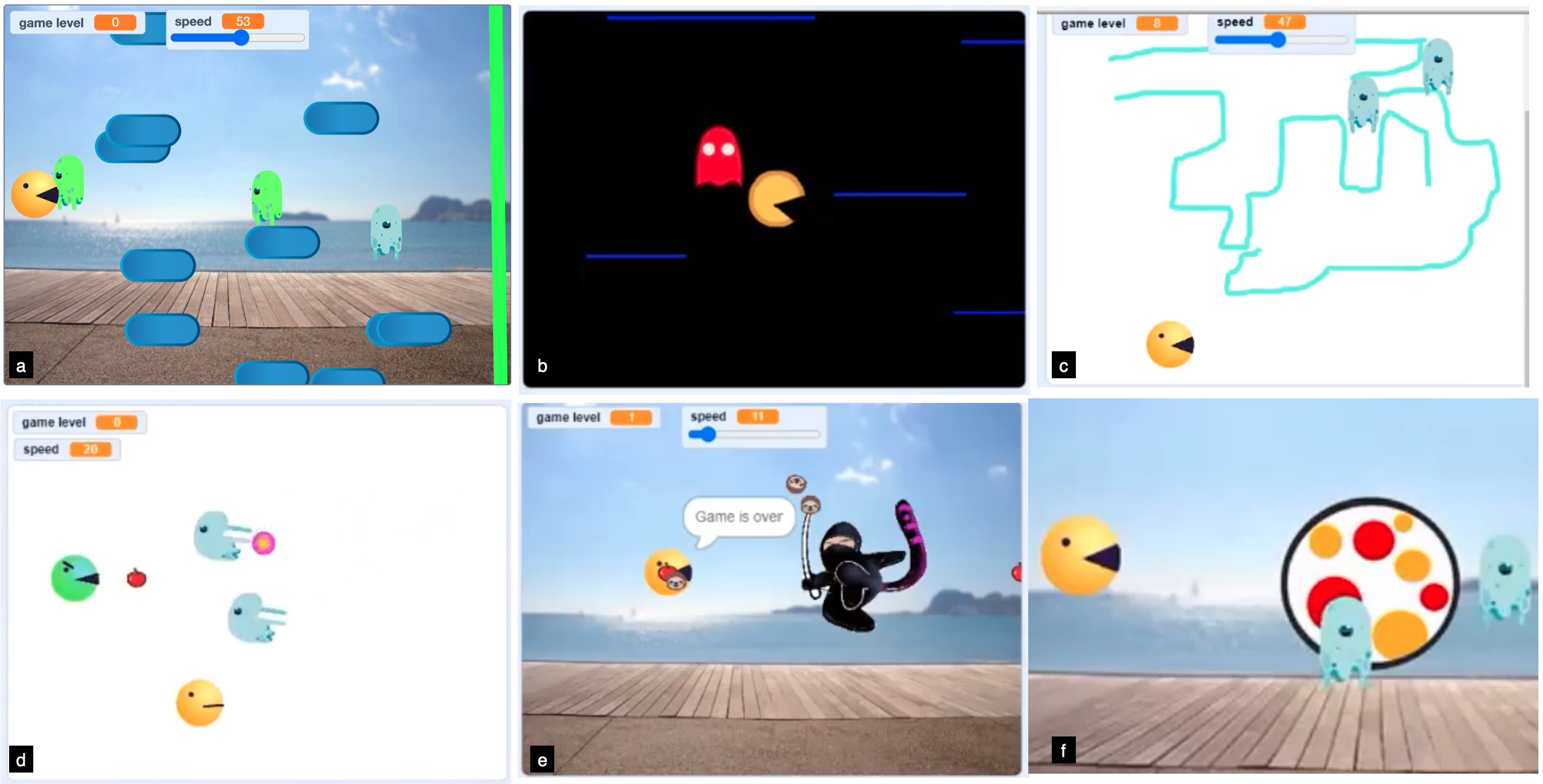}
\caption{Examples of family games from Session 1: (a) F5: game with multiple obstacles and animated disco ghosts, where Pacman needs to reach the green line to win; (b) F3: game with original artwork, with a dynamically changing maze and chasing ghosts; (c) F6: game with a hand-drawn maze and ghost clones; (d) F8: multiplayer game, where good Pacman competes against bad Pacman to shoot more ghosts; (e) F1: game with ninja shooting sloths, and (f) F2: game with a giant ball toppling ghosts on the screen.}
\label{fig:games_session_1}
\end{figure*}

\subsubsection{Session 1: Modifying a Game}
In the first session, children and parents worked with the AI Friend to modify the Pacman game. One game (from Family 5 (F5)) featured multiple obstacles and animated disco ghosts, where Pacman had to reach the green line to win. Another game (F3) had original artwork, a dynamically changing maze, and ghosts spawning at different parts of the maze to chase Pacman. In a third game (F6), Pacman had to navigate a hand-drawn maze followed by ghost clones. A fourth game (F8) was a multiplayer game with good and bad Pacman competing to shoot more ghosts. The fifth game (F1) involved Pacman fighting with a ninja that shoots at sloths. Finally, the sixth game (F2) featured a giant ball that could topple ghosts on the screen. Each game showcased unique features and challenges, and the AI Friend helped the families with coding, guidance, and idea generation (see game screenshots in Figure \ref{fig:games_session_1}).

This first session revealed that children's experiences with AI can vary considerably, depending on their individual preferences, prior experiences, and family dynamics. For example, collaborative coding with parents led to more positive outcomes (F1, F2, F3, F5, F7, F8), whereas sibling collaborations showed a dominance of older siblings, potentially leading to the exclusion of the younger sibling (F4, F6). In addition, sessions involving clear communication and guidance led children to accept AI Friend suggestions, resulting in improved games. However, negative experiences with the AI Friend, such as arguments between siblings or discontent with the AI's suggestions, could highlight the importance of individual family preferences and prior experiences when working with AI.

\begin{figure*}
\centering
\includegraphics[width=5.5in]{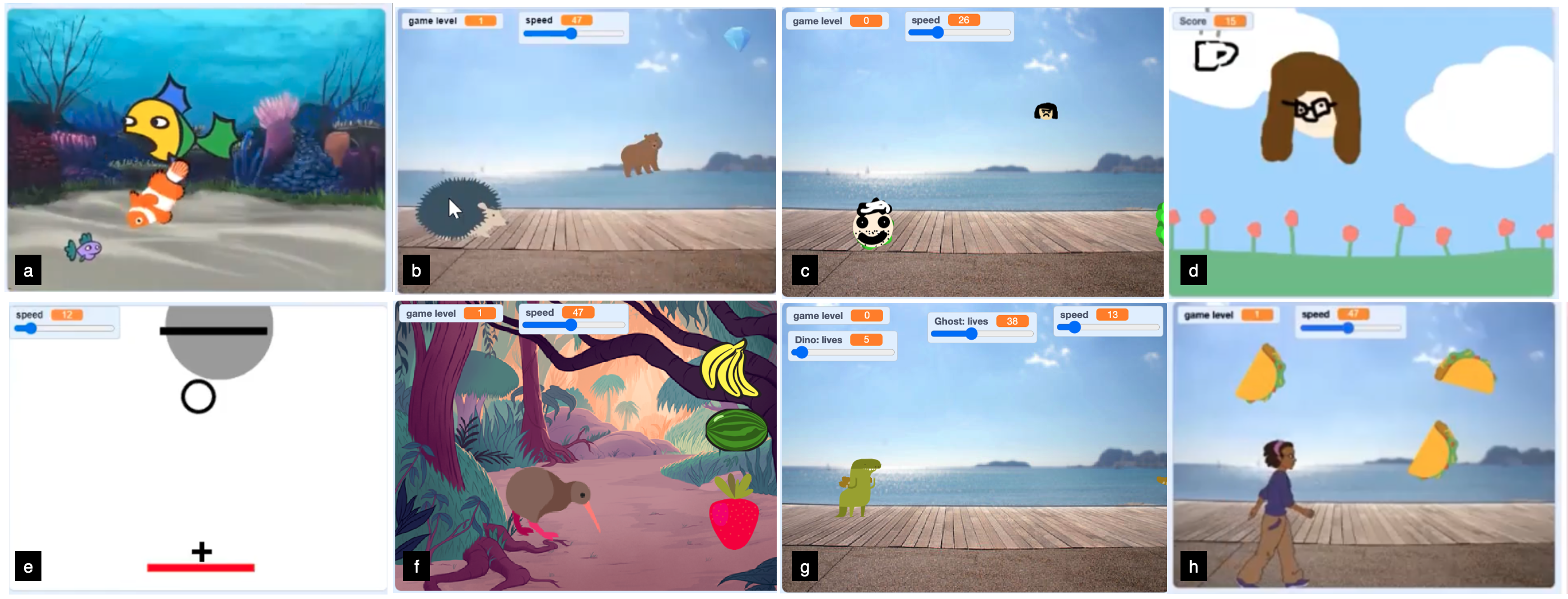}
\caption{Examples of family games from Session 2: (a) F1: a fish game, where a big fish is eating smaller fish and getting larger; (b) F6: a hedgehog stinging bears protecting gems; (c) F5: a child running into friends; (d) F4: a coffee drinking game; (e) F3: a ping pong game against a robot; (f) F2: a kiwi bird eating fruit; (g) F7: a dinosaur shooting bread at ghosts; and (h) F4: a girl dancing in a taco rain.}
\label{fig:games_session_2}
\end{figure*}

\subsubsection{Session 2: Making a Game from Patterns}
The second part of the study involved sessions where children and parents programmed a new game using programming patterns with the help of an AI Friend. Overall, families programmed various games using patterns, but also using AI support. For example, one game had hedgehog-stinging animated bears gliding over the screen to protect gems (F6). In another game, players could compete against a robot in a ping pong game, where the paddles created cool animations of the ball bouncing (F3). The game from F2 featured a kiwi bird eating fruit that gave it different points and F7 game featured a dinosaur who could shoot bread at ghosts and had the power to call the bread back. Each game was unique and demonstrated the creative ideas that families brought to their programming with the guidance of the AI Friend (see Figure \ref{fig:games_session_2}).

Overall, this session showed that the AI Friend prompts catalyzed collaborative coding, facilitated idea generation, and supported the development of more advanced coding skills with families. In addition, in this session, the AI showed the potential to support children's learning and creativity in coding various games, providing guidance, suggestions, and encouragement.

\begin{figure*}[t]
\centering
\includegraphics[width=6in]{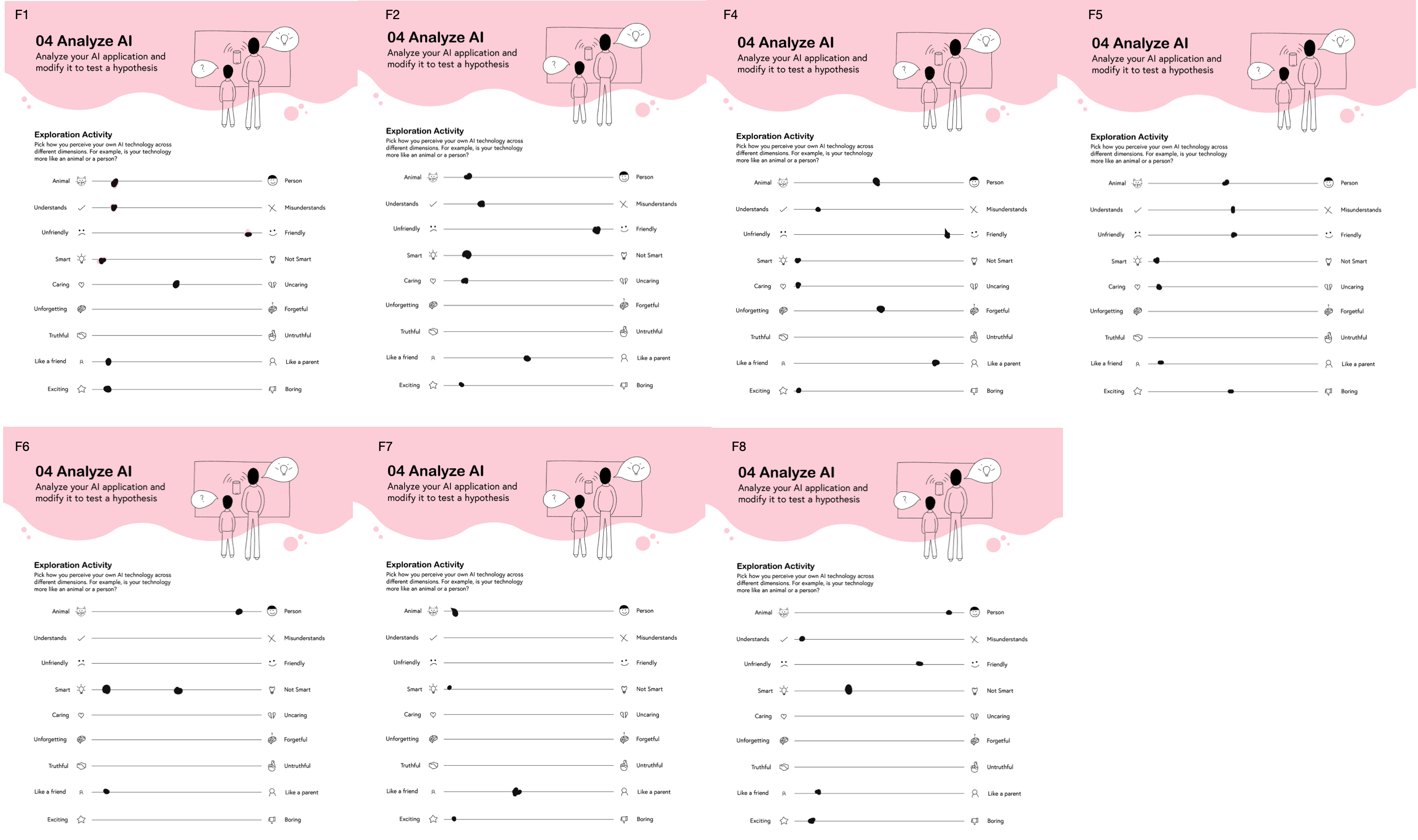}
\caption{Feedback from the AI Friend Persona Sheet and attributes from all families (F3 did not complete the sheet).}
\label{fig:ai_perception}
\end{figure*}

\subsubsection{Session 3: Final Interview}
The third study session involved interviews with children and parents asked to provide feedback about their experience interacting with the AI Friend. All children said they preferred the AI to ask questions or give hints to help them fix bugs or implement specific game behaviors rather than giving them the solution or fixing the programs. One child (F8) said they would like the AI Friend to show how code changes would trigger unusual behavior, while another (F1) said they would like the AI to ask clarifying questions when unsure how to help. All participants said the AI Friend's affirmations and encouragement helped a lot, and many children expressed a desire for the positive feedback to be personalized.

Several children noted how much they appreciated the AI Friend getting them unstuck and found it less frustrating than coding alone. However, they sometimes found the wording used by the AI to be confusing, and in those cases, they said images would be very helpful, especially for locating specific programming blocks. In addition, having a parent present helped them better understand the AI's suggestions and questions.
 
All children said they would like the AI Friend to be amiable and to understand them. They expressed a desire for the AI to be smart, but not too smart, and still let them figure things out by themselves (see Figure \ref{fig:ai_perception}). 

\subsection{Interactions with the AI Friend}
Our analysis revealed several distinct types of interactions with the AI Friend, each catalyzing different types of family collaboration, some positive, and some negative. (Each of these corresponds to a code in Table \ref{tab:aifriend_codes}; we do not report on all codes in the table due to length limits).
\subsubsection{Ideation with AI Friend}
\begin{figure*}
\centering
\includegraphics[width=5in]{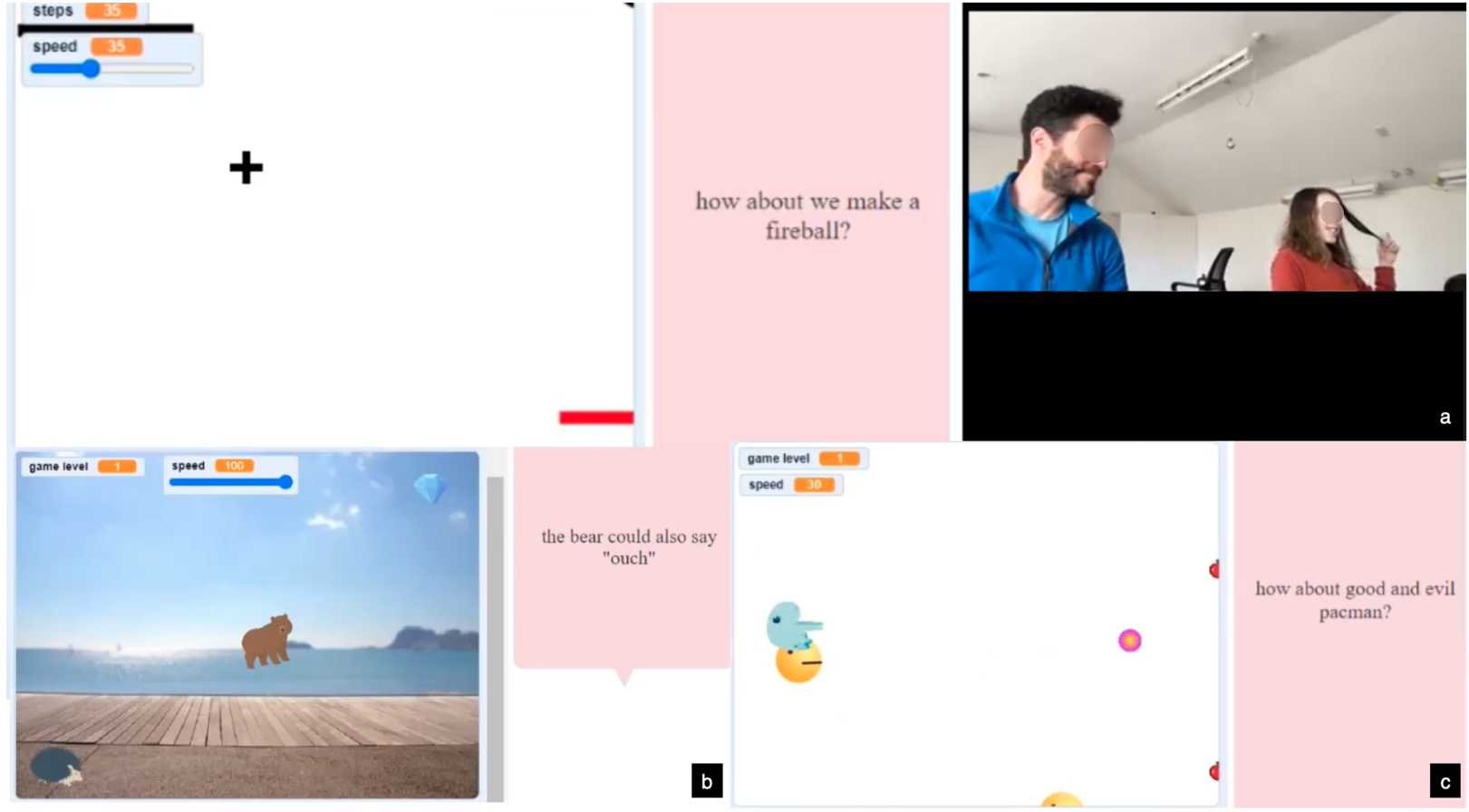}
\caption{Examples of AI Friend ideas: (a) F3: adding effects to the ping pong ball to make it look like a fireball; (b) F6: making the bear say ``ouch'' when touching the hedgehog; and (c) F8: creating good and evil Pacman characters for the multi-player game version.}
\label{ai_ideas}
\end{figure*}
The AI Friend's role in the game design process involved \textbf{stimulating and supporting ideation among families} (see Figure \ref{ai_ideas}). This was achieved by \textbf{asking questions} that guided families to choose and express their creative ideas. For example, the AI Friend of F3 suggested what would happen when a ball touched a paddle and suggested what to add to make the game more exciting. In response, the child expressed interest in adding the suggested effect and asked the AI Friend to help him implement it:
\begin{quote}
    \textit{``Should we add an effect on the ball when it touches the paddle?''} --- AI Friend's suggestions to the F3 family. \\
    \textit{``I should do that, but can you help me do that?''} --- G., 11 years old, responds to an AI Friend's suggestion. \\
    \textit{``How about we make a fireball?''} --- AI Friend responds (see Figure \ref{ai_ideas}a). 
\end{quote}

Moreover, the AI guided children struggling with generating ideas or experiencing a mental block. For instance, when a child (F4) was unsure of what should happen next in the game, the AI Friend suggested adding an effect and a motion direction, which guided the child to think further about the game. In this way, the AI served as a mentor that stimulated the child's creativity and helped them overcome mental block:
\begin{quote}
    \textit{``Hmm. It looks interesting. Just flying Tacos in the sky. Okay, how about you make it go down?''} --- AI Friend suggestions to F4 family. \\
    \textit{``Can it then go this way (points down the screen), and then what will happen?''} --- M., 10 years old, responds to AI suggestions. \\
    \textit{``Yeah. So when the space bar is clicked, we want to move it. How should we control the motion direction?''} --- AI responds. 
\end{quote}
In these two examples of ideation with the AI Friend (F3, F4), we see how the AI Friend encourages families to think creatively by \textbf{helping them generate, develop, and express their ideas.}

Additionally, the AI-supported family game development by \textbf{offering examples of game mechanics or elements that they could use} as a starting point to create their own unique game. For example, the AI Friend suggested that F8 create a good and evil Pacman, which gave the family an idea about how to modify their game (see Figure \ref{ai_ideas}a). 
The AI helped families develop their game designs by providing specific character examples and ideas and inspired them to create unique and engaging games.

S., a 12-year-old from family F8, expressed that he preferred the AI Friend's support in conceptualizing game ideas since it let him build upon the initial idea and use his creativity. He suggested that a separate section for art support would also be useful when he needs assistance in designing game artifacts:
\begin{quote}
    \textit{``I prefer help with game concepts because then you can mix them to build upon it. So it [referring to the AI Friend] gives you an idea, and then you can use your mind and creativity to do it, but maybe there could be another section where it focuses on the art and say, ``I know what to do with the cone, but how should I make everything look?''} --- S., age 12, F8
\end{quote}

Several children recognized that the AI Friend's ideas helped them when they did not know how to start their game or when the game was becoming boring. For example, H., a 12-year-old from family F2, acknowledged the usefulness of the AI Friend in \textbf{overcoming the challenge of starting a coding project and when encountering a roadblock in the creative process.} She believed that many people would appreciate the AI Friend's support in these situations:
\begin{quote}
    \textit{``Most people would like coding with AI Friends because one of the hardest parts of our project is when you start and also when you run into a wall, and you're out of ideas.''} --- H., age 12, F2 
\end{quote}

G., an 11-year-old from family F3, noted the helpfulness of the AI Friend in adding additional features and effects to their project, which added a fun element to their coding experience. They acknowledged that coding could become boring, and the AI Friend helped to prevent that. The child appreciated the support provided by the AI during their coding project:
\begin{quote}
    \textit{``The AI Friend was definitely helpful because I would not have had the funny speed thing and the effects if it wasn't there, it's just nice having that little extra bit of help during my scratch project because it does get pretty boring as my brain gets foggy.''} --- G., age 11, F3
\end{quote}

S., a 12-year-old child from family F8, expressed their belief that the AI Friend can play a valuable role in helping them develop their ideas over time. He believed that, with assistance from the AI Friend, he would eventually become self-sufficient and able to generate new ideas independently. This underscores the importance of the AI Friend as a \textbf{tool to foster independence and creativity} in children as they learn to program:
\begin{quote}
    \textit{``I think after a while you probably won't need (it) anymore because it has taught you enough. Maybe it can tell you like ``Oh, next time if you need more ideas, it can give you a way to think of new ideas, not just give you the ideas.''} --- S., age 12, F8
\end{quote}

M., a 10-year-old child from family F4, shared a similar outlook and expressed a \textbf{desire for the AI Friend to adapt its support over time}. She believed that even as she became more skilled in programming, the AI could continue to offer assistance by providing increasingly advanced information and guidance. This further emphasizes the importance of the AI being able to adjust its support to meet the changing needs of children as they grow and develop their programming skills:
\begin{quote}
    \textit{``Well, maybe even though you're really good, you'll still not understand something. So maybe the AI is like when you're younger, then it kind of just tells you the thing you need to know. And then when you get older, it tells you more about what you're doing.''} --- M., age 10, F4
\end{quote}

Our study reveals that AI Friends have the potential to be long-term partners in aiding children's learning and growth through game programming. They can foster independence while also adapting to their evolving needs. Furthermore, the AI's capacity to ask questions and offer advice is essential for stimulating ideation and creativity when families program games. It encourages children to communicate their ideas and conquer mental blocks and provides examples to motivate them to create unique and engaging games.

\begin{figure*}
\centering
\includegraphics[width=5in]{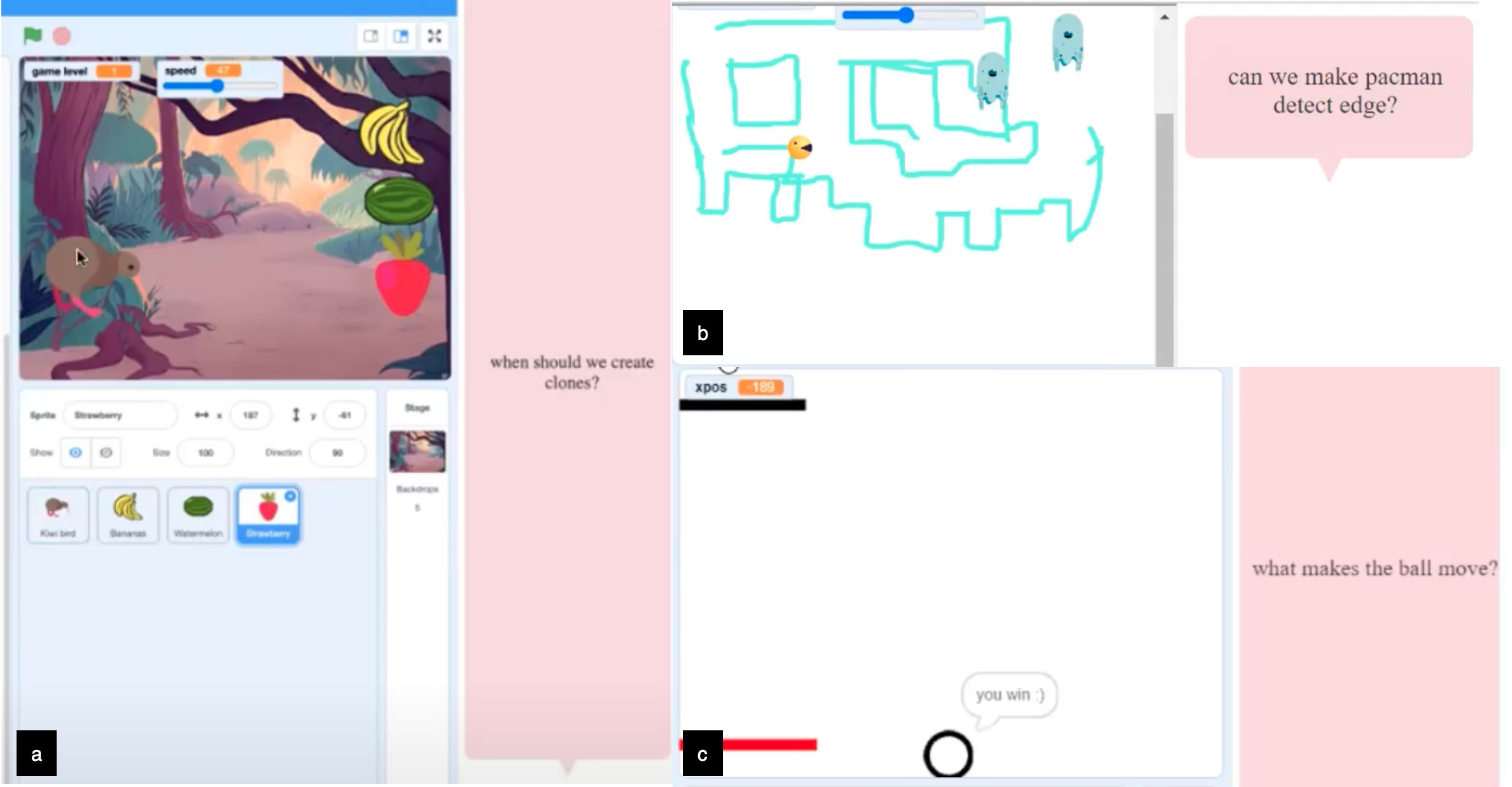}
\caption{Examples of AI Friend's support with debugging: (a) F2: AI Friend helps with creating clones for fruits; (b) F6: AI Friend assists with making Pacman avoid the maze; and (c) F3: AI Friend prompts child to think about triggering ball motion.}
\label{fig:ai_debugging}
\end{figure*}

\subsubsection{Debugging with AI Friend}

While the AI Friend catalyzed ideation, it also served as a debugging aid. It supported them in various ways, such as by fixing scripts, explaining how code works, encouraging specific tests, and asking logic questions (see Figure \ref{fig:ai_debugging}).

M., a 10-year-old from family F4, highlighted the importance of the AI Friend's understanding of the coding project. She suggested that it would make the coding experience more enjoyable if the AI could anticipate and correct mistakes before they occur, demonstrating a need for the AI to understand the context and goals of the coding project to provide more effective support: 

\begin{quote}
    \textit{``Which maybe having like never have it know what you're doing so that it'll get the idea? Maybe you don't have to ask him if you do something wrong. So like, I know that if I make something, and then I get something wrong, I get frustrated. But if they know what they're doing, if they know what you're doing, then they can correct you before that happens.''} --- M., age 10, F4
\end{quote}

Children also spoke about the value of the AI Friend in explaining code when it became confusing. They noted that it could be difficult to understand why the code is not working as intended, and the AI was most useful in these situations; they emphasized the need for the AI to provide clear and concise explanations of code or help them understand how the code executes: 

\begin{quote}
    \textit{``Sometimes it can be confusing when we write a lot of code and then run it. Sometimes we write a single block of code and run it to see if it works, but if it doesn't work, we don't always know exactly why.''} --- S., age 12, F8.\\

     \textit{``That was helpful [the AI Friend], definitely when explaining the show and hide thing that helped me a lot because I was confused about that. And it definitely helped me with it, going through it and understanding it more. I liked asking questions like: ``how are the blocks executed?'' because that made me realize I should try to find the answer more.''} --- G., age 7, F1
\end{quote}

One child, S., a 12-year-old from family F8, noted that the AI Friend could reduce frustration when encountering bugs in their code. With its help, S. could resolve coding issues more quickly than before, though S. also expressed concern about the potential for people to rely too much on the AI and not learn to debug code themselves. To address this, S. proposed that the AI be programmed to identify when a person is genuinely stuck versus when they are simply relying on the AI to do the work for them:

\begin{quote}
    \textit{``If you had put in a lot of effort, saying ``Oh, I worked so hard on this project, I spent countless hours,'' and someone else had just let the AI do the whole thing, you would feel like ``Why did I have to put in so much effort?'' Eventually, people will start relying too much on AI and not do it themselves..''} --- S., age 12, F8\\
    \textit{``It's AI, it can teach itself when the person is truly stuck and when they're just saying ``yes'' they are stuck..''} --- S., age 12, F8
\end{quote}

A., a 12-year-old from family F4, acknowledged that while the AI Friend helped debug her code, she did not want it to complete the entire project for her. Instead, she wanted to maintain the feeling of ownership over her game and not rely too heavily on the AI:  

\begin{quote}
    \textit{``Yeah, if it [AI Friend] does everything for you, it wouldn't really be your game at that point.''} --- A., age 12, F4
\end{quote}

M., a 10-year-old from family F4, shared that she found it helpful when the AI Friend explained new concepts while debugging, showing her how to use the ``broadcast'' feature to make objects disappear from the screen. She also compared her experience of debugging code with the AI versus her dad and noted that the former's suggestions were similar to what her dad had advised. This highlights the potential of the AI to complement parent support and provide additional resources for children as they learn to program: 
\begin{quote}
    \textit{``So, for making them disappear, you could use something like broadcast. So, if a girl touches a taco, then you would broadcast the message ``eaten.'' And then, on the taco sprite, you say, ``When I receive the message ``eaten'', hide.'' So, that's how you could make them[tacos] disappear.''} ---  M., a 10-year-old, F4\\
    \textit{``The AI Friend was good. Like, when they told me to say, ``Separate the scripts,'' that's exactly what my dad told me to do.'' --- G., age 11, F3}
\end{quote}
In family F7, 10-year-old M. wanted the AI Friend to provide guidance and support as they debug the code. He believed the AI should show him the next steps and help him identify the problem but still allow him to fix the issue by himself, which can be a rewarding experience. M. also emphasized the importance of AI not doing everything for children since this would detract from the learning experience: 
\begin{quote}
    \textit{``You can tell the AI ``I tried my best to show me the next step.'' It's rewarding when you fix it yourself..''} ---M. age 10, F7\\
    \textit{``It's good if the AI isn't doing everything for them [other kids], and it's showing them how they can fix it and search for it on the web. It would be really cool because then they'd actually be learning and not just cheating..'' --- M., age 10, F7, added later when reflecting on how the AI could help other children.}
\end{quote}

The comments made by M. (F7) resonated with the majority of reflections from the other children, who mainly wanted an AI Friend to help them help themselves. These findings highlight the potential of the AI Friend to serve as a supportive partner in helping families debug their code. Providing guidance and support while allowing the children to take an active role in fixing their issues, the AI Friend can enhance the learning experience and help the children develop important problem-solving skills.

\begin{figure}
\centering
\includegraphics[width=3.5in]{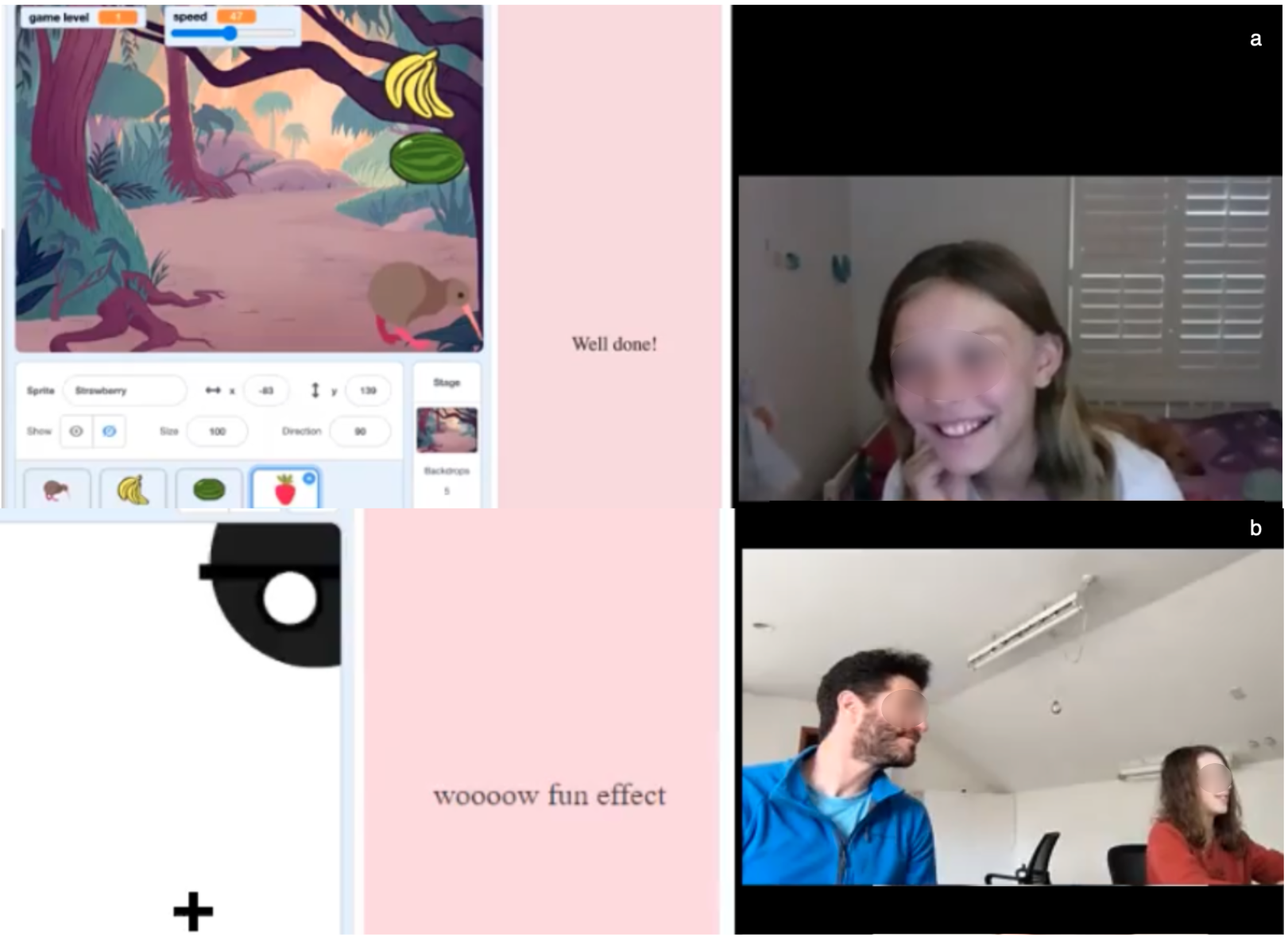}
\caption{Examples of affirmations provided by the AI Friend during the programming sessions: (a) F2: AI Friend praises H. for successfully getting her bird to eat fruits, and (b) F3: AI Friend congratulates the family for adding a fun effect to the ping pong ball.}
\label{fig:ai_affirmation}
\end{figure}

\subsubsection{Supporting Creative Coding Identity}

The AI Friend provided the encouragement and affirmations in Table \ref{tab:aifriend_prompts} to families during programming sessions, and these played an important role in their experience. Affirmations included praising a child for successfully getting a bird to eat fruits or congratulating the family for adding a fun effect to a ping pong ball (see Figure \ref{fig:ai_affirmation}), which helped to build confidence and motivation in the children and families.

For example, S., a 12-year-old child from family F8, emphasized the value of the AI Friend's affirmations, particularly during frustration while coding. Another child, H., age 12, F2, said that AI encouragement helped her finish her game. Children appreciated the positive reinforcement that the AI provided, which helped them to feel good about their accomplishments and maintain motivation:
\begin{quote}
   \textit{``Well, I like that because, sometimes, when you code, it gets frustrating when we finally get to work, it's good to let you feel good, and it's good to have someone say a good job''} --- S., age 12, F8\\
   \textit{``I really like receiving just like ``well done'' because it's it's like I'm being congratulated for work that I would not have been congratulated if it wasn't for the AI Friend; I needed that little incorrect encouragement to finish the project.''} ---H., age 12, F2
\end{quote}

The children also appreciated the AI Friend's personal touch through its affirmations. They felt that the AI's personalized encouragement, such as congratulating them for making a specific fix, was more meaningful than generic praise. 

Overall, the families appreciated the AI's role in providing encouragement and affirmations throughout their programming experience. These affirmations helped to build their confidence, foster a sense of accomplishment, and motivate them to continue working on their projects. The families even imagined the possibility of the AI being acknowledged in the credits of their projects, highlighting the significance of its role in their experience:

\begin{quote}
   \textit{``If you make this huge favorite project on Scratch and then in the credits section, it's just like this AI Friend mentioned for inspiration and stuff. That would be really funny.''} ---G., age 11, F3
\end{quote}

\subsubsection{AI failure}
Sometimes the AI Friend failed to effectively support families. In the case of family F5, confusion arose when the child attempted to implement a speed variable for the ghost character but discovered it was controlling Pacman instead. This experience highlights the importance of clear variable names in starter games to prevent confusion and facilitate effective guidance from the AI. Despite initial difficulties, the child was able to find the ``move'' block with the assistance of their mother and continue with the programming session:

\begin{quote}
    \textit{``It has to say speed but also do speed?''} ---  S., age 10, F5, understands why her ghost is not moving but does not know how to fix it.\\
    \textit{``What do you need it to do?''} --- Mom, F5, helping her daughter reason about what to do next.\\
    \textit{``I need to set it to move.''} ---  S., age 10, F5, responds to her mom while starting to look for the ``move'' block.
\end{quote}

In another instance in F8, the child stopped using the AI Friend to focus on their ideas and a conversation with their mother. The AI's suggestions were perceived as distracting since the child's game intent was not clear. However, once the child had established a clear idea for their game, they re-engaged with the AI, prompted by their mother, and built upon the AI's suggestions to improve the game.

In F4, a situation arose in which the two daughters were arguing, and the father was not mediating. During the coding session with her sister and father, M. described the AI Friend as not being helpful but not causing any harm. However, in a subsequent session where M. programmed alone, she described the AI as helping provide suggestions and correct her when necessary. This highlights the importance of creating supportive  environments conducive for children to effectively engage with AI:

\begin{quote}
    \textit{``It was okay. I mean, it didn't help, but it didn't do anything bad either.''} --- M., age 10, F5\\
    \textit{``I liked the AI Friend when he helped me with making the angry snore sound. It's like if I have a part that I don't really understand, and then I ask it about it, and maybe it'll tell me which part. And then, if I get something wrong, it may even correct me.''} --- M., age 10, F5, describing her interaction with the AI Friend when programming alone.
\end{quote}

These instances demonstrate that while the AI Friend can be a valuable tool in helping children program games, there may be instances where it fails to provide adequate support. This highlights the need for ongoing evaluation and improvement of AIs to ensure they effectively serve families' needs.

\subsubsection{Joint Engagement Support}

\begin{figure*}
\centering
\includegraphics[width=5.5in]{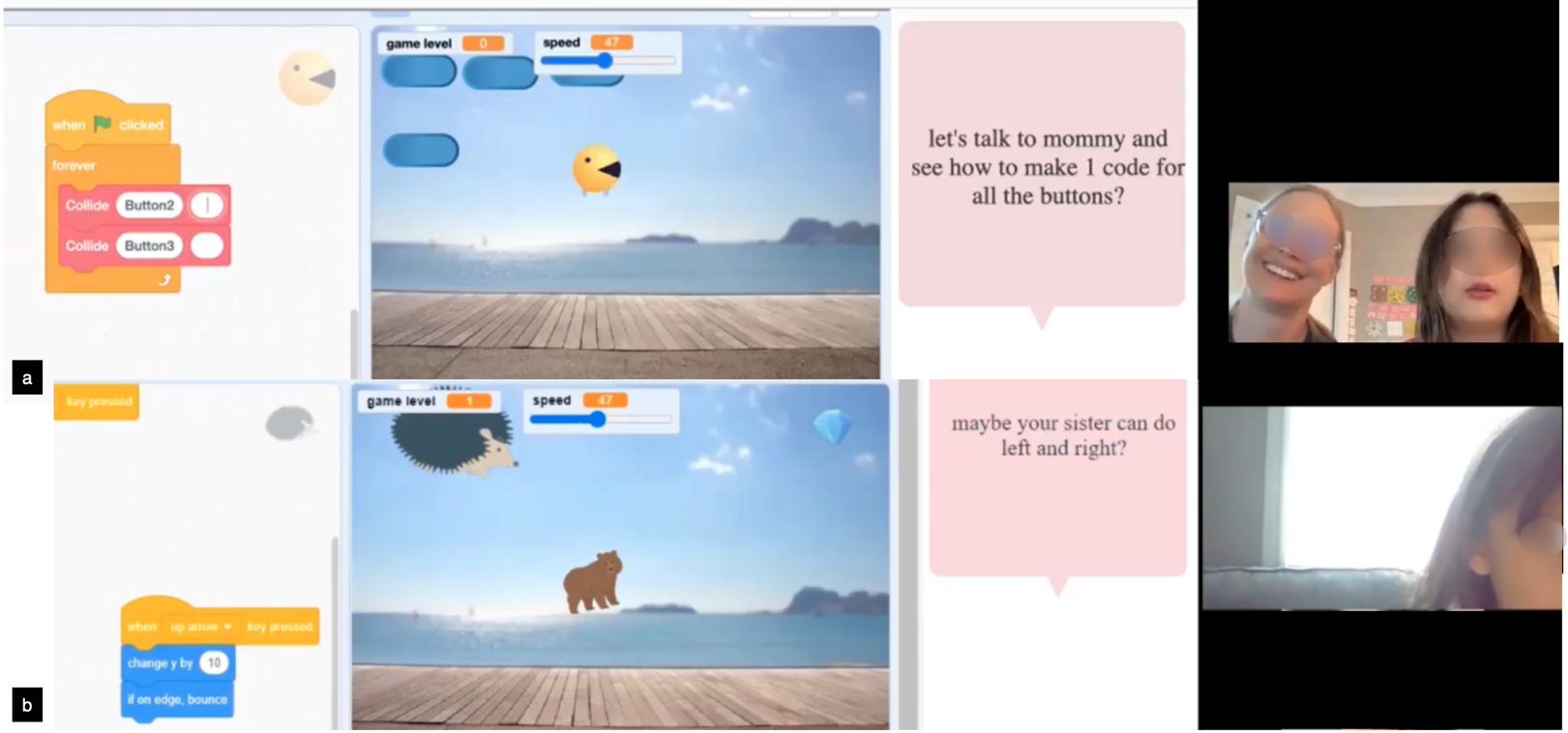}
\caption{Examples of the AI Friend encouraging joint family-AI engagement: (a) F5: prompting the child to talk to a parent about the "clone" concept; and (b) F6: prompting siblings to take turns coding.}
\label{fig:ai_support_joint_interaction}
\end{figure*}
Several patterns of joint family engagement emerged when children and parents collaborated with AIs. Mothers, fathers, and siblings supported each other in resolving technical challenges for study set-up, collaborating in the ideation process, debugging programs, interacting with the AI Friend, and brainstorming during co-design. Siblings primarily helped with ideation, programming, and co-design brainstorming (see Figure \ref{fig:fig_joint_interaction}).

During joint family programming, parents were particularly helpful when children did not understand the AI's suggestions. For example, J., a father from F1, assisted his son by asking questions and making suggestions regarding modifying the shooting programming pattern in their game. This interaction highlights the importance of parental support in fostering children's programming skills and helping them overcome challenges:
\begin{quote}
\textit{``So I have the Pacman here and a ninja, so if you press the space bar, it shoots the apple. Once you press the space bar, I want to create another sprite for the ninja.''} --- J., dad, F1, when helping his son modify the shooting programming pattern in his game.\\
\textit{``But what about on the side? How do we put it on the side? So I won't show it, just hide it.''} --- G., 7 years, F1, replying to his dad.\\
\textit{``I didn't know, so we have to make another one [code condition] when it touches Pacman or when it hits pigment. So how would we do that?''} --- J., dad, F1, replying to his son.
\end{quote}

Similarly, C., a mother from F5, helped her daughter reason about code duplication versus creating character clones by asking questions and guiding her thought process. This interaction highlights the importance of parents in facilitating children's critical thinking and problem-solving skills:
\begin{quote}
    \textit{``Do you want your code to do something different or the same?''} --- C., mom, F5, helping her daughter reason about code duplicates vs. code clones.
\end{quote}

In some cases, children relied on their parents to help them formulate questions for the AI Friend when seeking specific help. For example, S., a 10-year-old child from F5, asked her mother to help her express her question to the AI. This highlights the role of parents in supporting children's communication and collaboration skills:
\begin{quote}
    \textit{``How do I say what I am trying to do?''} --- S., age 10, F5, asking her mom to help her formulate a question for the AI Friend.
    \textit{``You could say: ``I need to figure out how to change speed''.''} --- Mom, F5, responding to her daughter's question.
\end{quote}

Parents also helped children explore the platform interface and test the different buttons and sliders to see how it would affect the AI's behavior: 
\begin{quote}
    \textit{``What do you think? What are you looking at? The icons at the bottom? Do you know what I see? `Help with code.' Maybe you should go back and see the answer to your questions from the robot.''} --- D., dad F4, helping his daughters find the answer from the AI. 
\end{quote}

M., a father from family F7, helped his son find answers to his questions by guiding him to follow the AI's suggestions. This interaction highlights the importance of parental support in helping children navigate and make use of the AI's resources: 
\begin{quote}
    \textit{``Did you see the AI suggestion to change the sprite name?''} --- M., dad F7, helping his son. 
\end{quote}

The AI also encouraged joint engagement between children and parents at times. For example, in F5, the AI prompted the child to talk to their parent about the concept of ``clones.'' In F6, the AI prompted the siblings to take turns coding, encouraging collaboration and teamwork (see Figure \ref{fig:ai_support_joint_interaction}).

Overall, the study revealed parents' crucial role in supporting children's programming skills and engaging with AI Friends. Parents helped children develop their critical thinking, problem-solving, communication, and collaboration skills through joint engagement. 

\subsection{Family Feedback on the AI Friend}

During the final interviews with families, several themes emerged regarding the design and functionality of the AI Friend. For example, one child, S., expressed a preference for interacting with the AI by both speaking and typing, recognizing the benefits of each mode of communication. Another child, M., discussed their desire for the AI to be less intrusive and to communicate through non-verbal cues, such as a smiley face, when they were focused on other tasks:

\begin{quote}
    \textit{``Talking is nice because you don't actually have to type anything. It feels real like you are actually in a conversation with someone. Typing is useful just in case it needs it or if it's easier to convey ideas in that way.''} --- S., age 12, F8\\
    \textit{``Maybe not always give me text the like maybe if I stop talking to it because I'm working on something and then I'll have that question for later. Maybe it'll type a smiley face when I want to talk to it.''} --- M., age 10, F5
    \textit{``I prefer to speak because like I feel like when I'm speaking like I can get the answer out more instead of figuring out what to type.''} --- H., age 12, F3
\end{quote}

Another child, G., emphasized the importance of giving children agency in customizing their collaboration with the AI Friend, suggesting that individuals should be able to personalize their AI to match their need for support. A related theme was the idea that the AI Friend should be able to tailor its support based on the child's preferences and previous interactions: 
\begin{quote}
    \textit{``Maybe for each person, they could like personalize their own bot. So if they don't want as much help, they can make the bot dumber. Have sliders so people can customize it. Some people need more help, and some people are there just for the ideas.''} --- G., age 11, F3
\end{quote}

Several children discussed the importance of being able to rate the AI's suggestions and feedback so that it could learn to better support them in their coding projects. One child suggested that the AI Friend's avatar could be customized based on the child's interests and preferences; another child emphasized the importance of human imagination and creativity, highlighting that even with extensive training, the AI would never be able to fully replicate the unique ideas and perspectives of individual children:

\begin{quote}
    \textit{``I think sliders are good because sometimes it's not just like, "Oh, this is completely bad." This is something that's in between, but I also want some text because it's not always just yes or no questions. You can say, ``Oh, I like that idea.'' ''} --- S., age 12, F8
    \textit{``I want to have a way to rate each AI Friend message as more helpful or less helpful, so it learns how to help me.''} --- M., age 10, F7
    \textit{``If you don't really want to do that idea, then you can do the thumbs down and ask ``Can you give me another idea?'' ''} --- H., age 12, F2
\end{quote}

Finally, one child suggested that it would be valuable to test the AI Friend in a school setting to see how other children respond to its assistance and that having another person to help them program could be more beneficial than relying solely on a teacher:
\begin{quote}
    \textit{``It would be great to test an AI Friend in schools and see how other kids like it when they code on it. Having another person helping them is better than just the teacher alone.''} --- G., age 11, F3
\end{quote}
This feedback highlights the children's perspectives on what makes a successful AI Friend in the context of programming games and underscores the importance of designing AI systems that are flexible, responsive, and tailored to the individual needs and preferences of children.

\section{Discussion}

Our work asked, \textit{How might children and parents engage in collaborative creative coding supported by an AI Friend?}
Our study revealed that our AI Friend's prompts and responses facilitated families' collaborative creative coding by helping to generate and express game ideas, support game debugging, and elaborate on family members' ideas, as well as by cultivating children's creative self-efficacy.
Our observations also revealed that the AI Friend did not do this alone: parents played a key role during the sessions by aiding children's programming skills and scaffolding the interaction with the AI when necessary. We also found that a family's interpersonal dynamics appeared to play a significant role in shaping what kinds of experiences had with the AI Friend: in some cases, it brought families together, appearing to heighten creative self-efficacy (CSE), and in other cases, it was a catalyst for conflict.

These findings are consistent with prior work on Joint-Media Engagement, where peer, parent, and teacher-supported behavior and classroom atmosphere emerge as significant factors in the process of development of CSE \cite{Beghetto2006-ex, karwowski2015development}. Similar to findings from Tang et al., who studied STEAM creative ideation~\cite{tang2022parents}, we found that parental engagement supported student creative self-efficacy in creative coding by resolving technical challenges for study set-up, collaborating in the ideation process, debugging programs, interacting with the AI, and brainstorming during co-design. Parents helped children develop their creative coding, problem-solving, communication, and collaboration skills through joint engagement. Our results extend this work, demonstrating that intelligent support can participate in family media creation in ways that amplify family interactions rather than disrupt them.

Our findings also mirror emerging evidence about professional software developers' experiences with AI supports such as GitHub Copilot. In one study, for example, participants preferred to use Copilot in daily programming tasks since it often provided a useful starting point and saved the effort of searching online. However, participants faced difficulties in understanding, editing, and debugging code snippets generated by Copilot to various degrees based on their level of experience \cite{vaithilingam2022expectation}. In the same way, families' ability to interact productively with the AI Friend appeared to be mediated by their collective prior knowledge of programming and problem-solving, which shaped their capacity for interpreting its prompts.

Some limitations to internal and external validity limit the interpretation of our findings. First, and most obviously, the AI Friend was not ``real'' in that it was a researcher pretending to be an intelligent agent. An actual agent, enabled by technologies such as large-language models, would likely have behaved differently in subtle ways and may have led to different reactions from families. Second, it was not possible to systematically observe every child's interaction with every family member, and some children spoke less in different families; it may be that children who verbalized more reasoned differently than those who verbalized less. Third, for the interactions we \textit{could} observe, observing a child's reason about the AI does not necessarily indicate ground truth for their conceptions; for example, it may be the case that children were reasoning in similar ways but were verbalizing their reasoning differently. Finally, our study sessions did not cover the possible ways that culture, community, and collaboration might have shaped creative coding. Since our analysis was episodic rather than temporal, innovative coding strategies may have been highly variable within individual and family behavior. Therefore, while a modest interpretation of our results suggests that our system supported family creative coding in our particular intervention, other populations could reveal different behaviors.

In light of these limitations, and the broader prior work on Joint-Media Engagement and Creative Self-Efficacy, our findings suggest some promise in building intelligent creative coding supports for families and further examining their capacity to promote creative self-efficacy. However, they also suggest many design and engineering challenges before such interactions are feasible and equitable:

\begin{itemize}
    \item Children in our study appreciated when the AI generated the right questions to help them fix their code or implement a new game behavior, rather than simply giving them answers. This suggests that agents might need to support both incidental triggers (deduced from platform use) and voluntary triggers (questions given by family members), as well as ways of balancing which are provided based on context. This is currently an unsolved problem in many program synthesizers for novice programmers \cite{jayagopal2022exploring}. 
    \item Families preferred the chat-based interactions, but this has implications for how agents might provide code recommendations, suggesting the need for more iterative and dialogue-based program synthesis, such as that in Ghostwriter Chat \cite{Ghostwriter:online} and Socrates \cite{ross2023case}. But this also raises questions about privacy, surveillance, and identity, as such techniques require an interaction history to be stored that may not neatly map onto an individual or group.
    \item Families preferred voice interactions for their ability to enable prompts without interrupting their programming flow. But implementing this feature poses equity gaps, as children's speech and foreign accents are often poorly recognized \cite{kennedy2017child}.
\end{itemize}

In sum, our study demonstrates that AI-enabled programming assistants can be a fruitful addition to enabling families to develop creative coding self-efficacy. Future work on AI programming assistants, if it seeks to work for children, must attend to broad variation in prior knowledge, the multi-user nature of family collaboration, the unpredictable trajectories of creative coding projects, and children's evolving capacity to make use of agents as they learn. And it should likely do so \textit{with} families, not for them. This context and application-informed design of AI programming assistants stand in contrast to the highly decontextualized nature of current AI program assistants and could be a rich area to understand how to create more human-centered, collaborative experiences with AI agents in general. And it could point to more general ways to reconcile the rapid proliferation of AI programming assistants into computing education learning contexts in general.

\section{Conclusion}
By developing and leveraging children’s creative efficacy and imagination, we would allow them to be prepared for the 21st century and inspire and advance our use of computational tools in unique and unforeseeable ways, such as learning how to code by collaborating with AI assistants. As our world moves, so do our art and our creativity. As researchers and designers, we must decide how much we want to include intelligent technologies in our creative and learning tools and for what purposes. Engaging children and parents as design partners in our creative coding and future tools design will ensure a future worth building up through.

\bibliographystyle{ACM-Reference-Format}
\bibliography{uwthesis,refs_tc,refs_ge,refs_ai,refs_idc21,refs_ge2,refs_thesis,refs_citizen,refs}
\end{document}